\documentclass[twocolumn,aps,prl,superscriptaddress]{revtex4-1}
\usepackage{amssymb}
\usepackage{amsmath}
\usepackage[pdftex]{graphicx}
\usepackage{physics}
\usepackage{color}

\begin{document}

\title{Realizing Hopf Insulators in Dipolar Spin Systems}

\author{Thomas Schuster}
\affiliation{Department of Physics, University of California, Berkeley, California 94720 USA}
\author{Felix Flicker}
\affiliation{Department of Physics, University of California, Berkeley, California 94720 USA}
\affiliation{Rudolph Peierls Centre for Theoretical Physics, University of Oxford, Department of Physics, Clarendon Laboratory, Parks Road, Oxford, OX1 3PU, UK}
\author{Ming Li}
\affiliation{Department of Physics, Temple University, Philadelphia, Pennsylvania 19122, USA}
\author{Svetlana Kotochigova}
\affiliation{Department of Physics, Temple University, Philadelphia, Pennsylvania 19122, USA}
\author{Joel E. Moore}
\affiliation{Department of Physics, University of California, Berkeley, California 94720 USA}
\affiliation{Materials Science Division, Lawrence Berkeley National Laboratory, Berkeley, California 94720, USA}
\author{Jun Ye}
\affiliation{JILA, National Institute of Standards and Technology and Department of Physics,
University of Colorado, Boulder, CO 80309, USA}
\author{Norman Y. Yao}
\affiliation{Department of Physics, University of California, Berkeley, California 94720 USA}
\affiliation{Materials Science Division, Lawrence Berkeley National Laboratory, Berkeley, California 94720, USA}
\date{\today}

\renewcommand{\b}{\textbf}

\newcommand{\FigHopf}{
\begin{figure}
\centerline{{\includegraphics[width=\columnwidth]{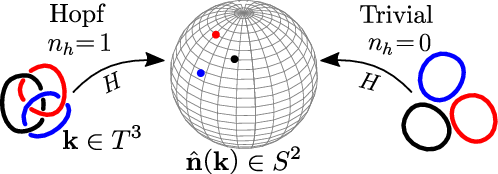}}}
\caption{
Three-dimensional two-band systems implement maps from the Brillouin zone $T^3$ to the Bloch sphere $S^2$. The pre-image of any point in $S^2$ is a closed loop in $T^3$. There exist topologically non-trivial states, Hopf Insulators (HIs), in which the pre-images of any two points on $S^2$ are \emph{linked} in $T^3$. HIs are characterized by a non-zero Hopf invariant $h$ equaling the linking number of the loops; pictured schematically are three points on $S^2$ and their pre-images in $T^3$ for both a HI with $h=1$ (left) and a trivial insulator with $h=0$ (right).
} 
\label{fig:Hopf}
\end{figure}
}

\newcommand{\FigEdgeStates}{
\begin{figure}
\centerline{{\includegraphics[width=\columnwidth]{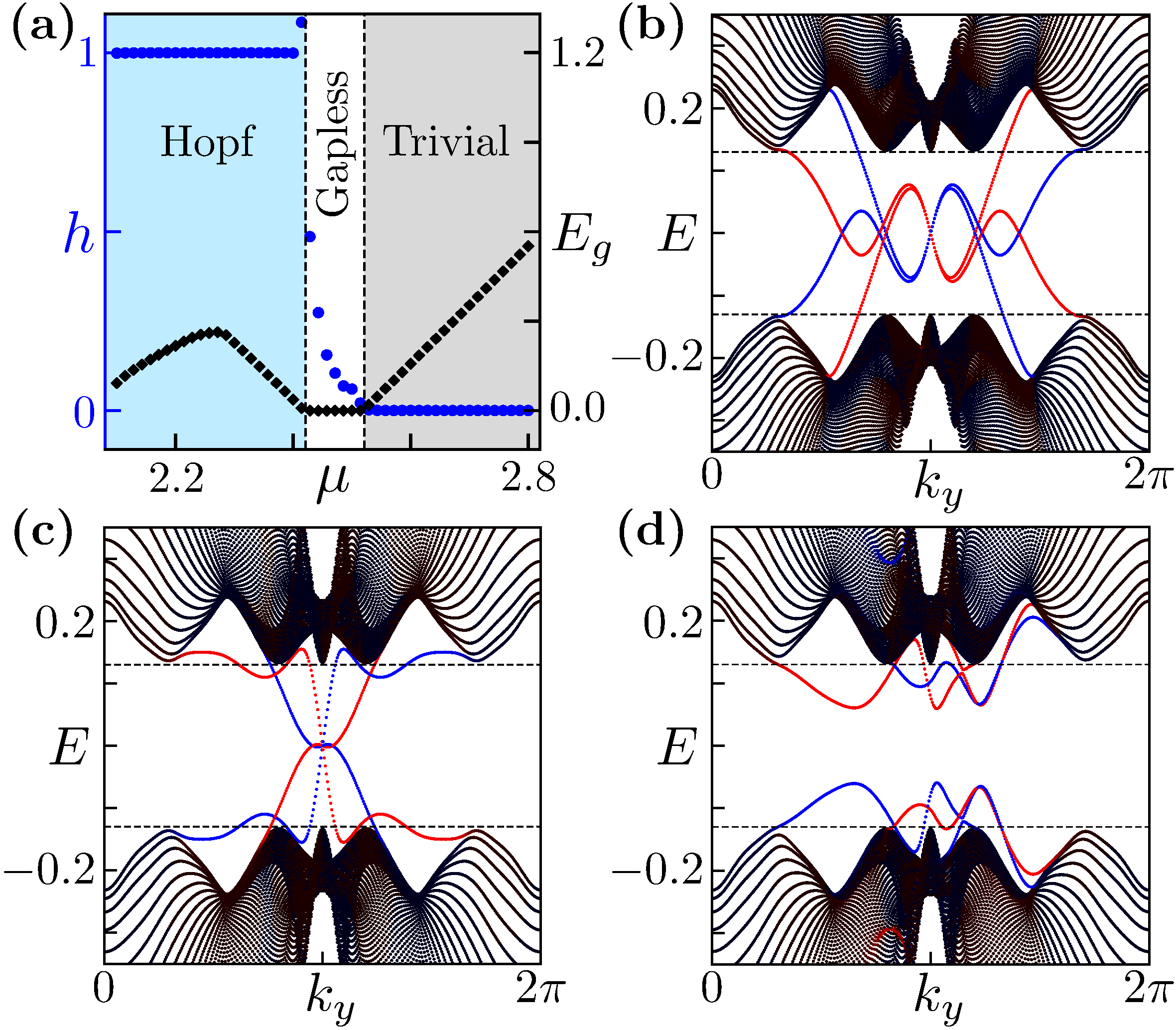}}}
\caption{
%{\color{red} To do: I will bring the letter labels outside of the graphs before resubmitting.}
$\textbf{(a)}$ Hopf invariant $h$ (left axis) and band gap $E_g$ (right axis) as a function of the staggered chemical potential $\mu = (\mu^A - \mu^B)/2$, found by discretizing the Floquet engineered dipolar Hamiltonian using $70\times70\times70$ $k$-points, periodic boundary conditions, and setting the nearest-neighbor inter-sublattice hopping in the $xy$-plane to $1$. The remaining plots show the energy spectrum with $\left(100\right)$ edge terminations. Black states indicate the bulk, red/blue indicate states localized to the left/right edge respectively, and dashed lines the bulk band gap. 
$\textbf{(b)}$ Adiabatic edge termination over 20 sites. The conducting edge states are protected by the $h=1$ topological invariant. 
$\textbf{(c)}$ Sharp edge termination respecting the $\mathcal{J}$ crystalline symmetry [Eq.~\eqref{eq:Cenke}]. The edge states are now protected by the symmetry. 
$\textbf{(d)}$ Introducing terms that break the $\mathcal{J}$ symmetry gaps the edge states. 
}
\label{fig:edge_states}
\end{figure}
}

\newcommand{\FigExpt}{
\begin{figure}
\centerline{{\includegraphics[width= 0.8\columnwidth]{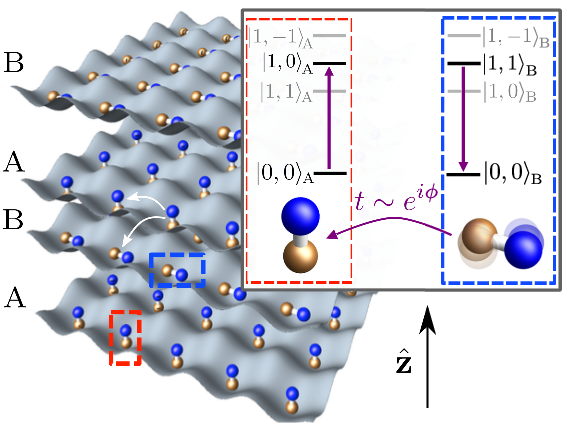}}}
\caption{
The proposed experimental set-up consists of dipolar molecules confined in a three-dimensional optical lattice, with two sub-lattices $A$ and $B$ separated in the $z$-direction. A combination of applied electric and magnetic fields and the intensities of the lattice beams themselves set the molecules' rotational axes along the $z$-direction, and are tuned so that $\ket{J=1,m=0}$ excitations (depicted as $z$-oriented molecules) on the $A$ sub-lattice can `hop' to $\ket{J=1,m=1}$ excitations (depicted as molecules spinning in the $xy$-plane) on the $B$ sub-lattice via the dipolar interaction, while conserving energy. 
Adding space- and time-dependence to these parameters leads to Floquet modulations $\mu^{\alpha}_{\b{v}}(t)$ of the on-site energies, allowing further control over the hopping magnitudes.
}
\label{fig:expt}
\end{figure}
}

\begin{abstract}
The Hopf insulator is a weak topological insulator characterized by an insulating bulk with conducting edge states protected by an integer-valued linking number invariant. The state exists in three-dimensional two-band models. We demonstrate that the Hopf insulator can be naturally realized in lattices of dipolar-interacting spins, where spin exchange plays the role of particle hopping. The long-ranged, anisotropic nature of the dipole-dipole interactions allows for the precise detail required in the momentum-space structure, while different spin orientations ensure the necessary structure of the complex phases of the hoppings. Our model features robust gapless edge states at both smooth edges, as well as sharp edges obeying a certain crystalline symmetry, despite the breakdown of the two-band picture at the latter. In a companion manuscript~\cite{schuster2021floquet}, we provide a specific experimental blueprint for implementing our proposal using ultracold polar molecules of  $^{40}$K$^{87}$Rb.
\end{abstract}

\maketitle

Topological insulators (TIs) exhibit conducting surface states protected by the existence of topological invariants associated with their underlying spin-orbit-coupled band structures~\cite{thouless1982quantized,haldane1988model,kane2005z,konig2007quantum,fu2007topological,moore2007topological,roy2009topological,zhang2009topological}. The past decade has seen a tremendous amount of progress in classifying and understanding the physical properties of these states. In particular, the interplay between a system's symmetries and dimensionality leads to a rich landscape of topological insulators, captured by the so-called `Tenfold Way' classification of free fermion theories~\cite{schnyder_classification_2008,kitaev_periodic_2009}. More recently, the structure of the table has been refined with the inclusion of crystal symmetries, giving topological crystalline insulators~\cite{fu_topological_2011} and higher-order topological insulators~\cite{schindler2018higher}. An ongoing task is to find physical realizations of further entries in the table; in cases where material implementations have not been found, the same topological states have often instead been realized in ultracold atomic gases trapped by optical lattices~\cite{goldman2016topological}.

Despite the ubiquity of the Tenfold Way, there are still topological states beyond its remit. One example is the Hopf Insulator (HI)~\cite{moore_topological_2008,deng_probe_2016,kennedy_topological_2016,liu_symmetry_2017}. Existing in three dimensions in the absence of any time-reversal or particle-hole symmetries, the Tenfold Way predicts that no topological state can exist, yet the Hopf Insulator features an insulating bulk and conducting edges protected by an integer-valued $\mathbb{Z}$ topological invariant. The apparent contradiction is avoided in two ways. First, HIs are weak TIs, meaning they only exist in two-band models, and the addition of further non-interacting bands can destroy the topology. Weak TIs are not captured by the Tenfold Way. Second, the $\mathbb{Z}$ topological invariant is not the usual Chern number, but is instead a \emph{linking number} familiar from knot theory and deriving from a relation to the Hopf map~\cite{hopf1931abbildungen,moore_topological_2008}. Despite a recent resurgence of interest in HIs~\cite{moore_topological_2008,deng_hopf_2013,deng_systematic_2014,deng_probe_2016,kennedy_homotopy_2015,kennedy_topological_2016,liu_symmetry_2017,ackerman2017static,wang_scheme_2017,tarnowski2017characterizing,yan_nodal-link_2017,alexandradinata2019actually,schuster2019floquet,unal2019hopf,he2019three,he2020non,hu2020quench}, fundamental difficulties have led to only a few proposals for their physical implementation~\cite{deng_probe_2016}. 

\FigHopf

There are three main barriers to implementing HIs in any tight-binding model of (say) electrons hopping on a lattice. First, the necessity of having precisely two bands rules out many material implementations. Second, as we will see, the nature of the non-zero linking number invariant requires a delicate structure in reciprocal space, meaning real-space interactions must be specified to large distances. Third, a strong spin-orbit coupling is required between the two bands.

In this Letter, we demonstrate that these barriers may be overcome by implementing HIs in lattices of dipolar interacting spins. Electron `hopping' can be replaced by transitions between rotational eigenstates, which are much easier to create and control; the long range of the dipole-dipole interaction then naturally realizes the long-distance hoppings. The two bands can be created from two sub-lattices, and different spin orientations can lead the hoppings to have a complex phase structure able to simulate a strong spin-orbit coupling. Further, we demonstrate that the key experimental signatures of HIs, gapless edge states, are present at any smooth (adiabatic) termination of our model and are robust to all smooth perturbations. This is in support of previous theoretical arguments for topologically-protected gapless modes at smooth boundaries, where translational-invariance and, as a consequence, the two-band picture are preserved. Nevertheless, we show that gapless edge states may persist at judiciously-chosen sharp (non-adiabatic) edges, owing to a crystalline symmetry that stabilizes the Hopf insulator to higher bands. This connects recent work predicting this `crystalline-symmetric Hopf insulator' with past numerical findings, which similarly observed gapless modes at sharp edges. 

\emph{Model.}---The Hopf Insulator (HI) can be understood by considering the three-dimensional two-band system at half filling described by the Hamiltonian
\begin{align}
\hat{H}(\mathbf{k}) = \sum^{3}_{i=1}n_{i}(\mathbf{k})\hat{\sigma}^{i}
\label{eq:Hamiltonian}
\end{align}
with Pauli matrices $\hat{\sigma}^{i}$. The bulk of the system is assumed gapped, requiring $|\mathbf{n}(\b{k})|>0$. Eq.~\eqref{eq:Hamiltonian} defines a map from the three-torus $T^3$ --- the Brillouin zone in which the wavevectors $\mathbf{k}$ reside --- to the Bloch sphere $S^2$ describing the possible states of the two-band system. The pre-image of any point on $S^2$ is then a closed loop in $T^3$. Restricting attention to the case in which the system has zero Chern number across any two-dimensional slice through the three-torus, the linking number of any two of these loops (necessarily an integer) is equal to the Hopf invariant of the map~\cite{hopf1931abbildungen}:
\begin{align}
h = -\frac{1}{4\pi}\int_{\textrm{BZ}}\textrm{d}^3\mathbf{k}\sum_{ijk}\epsilon^{ijk}A_{i}\partial_{j}A_{k}
\label{eq:Hopf}
\end{align}
with $\epsilon_{ijk}$ the Levi-Civita symbol, $\partial_i=\partial/\partial k^i$, and $A_{i}\left(\mathbf{k}\right)=-i\langle u_\mathbf{k}|\partial_i|u_\mathbf{k}\rangle$ the Berry connection for eigenstate $|u_\mathbf{k}\rangle$. Changing $h$ requires the gap to close. The $h=0$ state, in which all loops are unlinked, is a trivial insulator, and so for $n_{i}(\mathbf{k})$ such that $h>0$ the system is in a topologically non-trivial state: the Hopf insulator~\cite{moore2007topological,moore_topological_2008}. This $\mathbb{Z}$ topological invariant is fundamentally distinct from the Chern number appearing in the Tenfold Way. The situation is shown schematically in Fig.~\ref{fig:Hopf}. The HI is a weak TI, meaning that mixing with further non-interacting bands can destroy the topology. 

Recently~\cite{liu_symmetry_2017}, it was realized that if the system obeys a certain crystalline symmetry,
\begin{equation}
\label{eq:Cenke}
\mathcal{J}^{-1} \hat{H}\left(\mathbf{k}\right) \mathcal{J}  = -\hat{H}\left(\mathbf{k}\right)^{*},
\end{equation}
where $\mathcal{JJ}^* = -1$, the HI is promoted to a \emph{strong} TI characterized by a $\mathbb{Z}_2$ invariant (an example of a topological crystalline insulator, TCI)~\cite{liu_symmetry_2017}. In two-band models at half-filling this symmetry is always present, with any half-filled two-band Hamiltonian obeying Eq.~\eqref{eq:Cenke} with $\mathcal{J} = \hat{\sigma}_y$. In systems with more than two bands it can be viewed as the composition of inversion and particle-hole symmetries.

\FigExpt

Our implementation of the Hopf Insulator is based on the following Hamiltonian:
\begin{align}
\label{eq:Htb}
\hat{H}_{\textrm{eff}} = & \frac{1}{2}\sum_{\mathbf{v},\mathbf{r}\ne\mathbf{0}}\sum_{\alpha\beta} t^{\alpha \beta}_{\mathbf{r}} \hat{a}^{\dagger}_{\mathbf{v} + \mathbf{r}, \alpha} \hat{a}^{\phantom{\dagger}}_{\mathbf{v}, \beta} + \textit{H.c.}\nonumber\\
& + \sum_{\mathbf{v}}\sum_{\alpha} \mu^{\alpha}  \hat{a}^{\dagger}_{\mathbf{v}, \alpha} \hat{a}^{\phantom{\dagger}}_{\mathbf{v}, \alpha}
\end{align}
where $a^{\dagger}_{\mathbf{v}, \alpha}$ creates a hard-core boson at lattice site $\mathbf{v}$ and sub-lattice $\alpha\in\{A, B\}$. We adopt a bosonic rather than fermionic description, permitted by the single-particle nature of the HI, as we will model the hopping of electrons between sites with the exchange of angular momentum eigenstates. The sum over positions $\mathbf{r}$ indicates the presence of long-range hoppings necessary to realize the delicate $k$-space structure needed for all loop pre-images to link. The model has two sub-lattices, which will form the two bands. Both intra- and inter-sub-lattice hoppings are present, $t^{\alpha \beta}_{\mathbf{r}}$, as well as a sub-lattice-dependent chemical potential $\mu^{\alpha}$. 

\emph{Dipolar Hopf insulator.}---We propose an implementation of the Hamiltonian of Eq.~\eqref{eq:Htb}, which can naturally be realized in three dimensional lattices of dipolar interacting spins. 
Our proposal can be realized in a number of experimental platforms, ranging from highly-magnetic neutral atoms such Erbium and Dysprosium~\cite{lu2010trapping,aikawa2012bose,ferrier2018scissors,trautmann2018dipolar} to strongly-coupled solid-state spin defects~\cite{kucsko2018critical, cai2013large, barkeshli2015continuous} to Rydberg-dressed atom tweezer arrays~\cite{zeiher2017coherent,bernien2017probing,de2019observation}.
Here, we focus on ultra-cold polar molecules trapped in a three-dimensional optical lattice (Fig.~\ref{fig:expt}), where tunable strong dipolar interactions have already been experimentally demonstrated \cite{yan_observation_2013,hazzard2014many}.
Recent progress has led to the development of numerous molecular species for such set-ups~\cite{sage2005optical,ni2008high,park2015ultracold,takekoshi2014ultracold,guo2016creation,ciamei2018rbsr,molony2014creation,tung2013ultracold,deiglmayr2010permanent}. 
To demonstrate that this proposal is accessible in near-term experiments, we provide a detailed, quantitative blueprint for its implementation in the specific case of ${}^{40}$K$^{87}$Rb~\cite{ni2008high,moses2015creation,yan_observation_2013,ospelkaus_controlling_2010,aldegunde_manipulating_2009,aldegunde_hyperfine_2008} in a companion manuscript~\cite{schuster2021floquet}.

The basic geometry of the setup we envision is a three-dimensional optical lattice generated using four pairs of counter-propagating beams: two pairs forming the $xy$-lattice and two pairs forming the $A$ and $B$ sub-lattices in the $z$-direction (Fig.~\ref{fig:expt}). We assume the molecules completely fill the lattice, and each molecule is well-localized to its site by a deep confining potential. Rather than having molecules physically hop between lattice sites, we instead utilize the molecules' rotational degrees of freedom to simulate hard-core bosonic excitations. At lowest order, these rotational states are governed by the Hamiltonian $\hat{H}_{\text{rot}} = \Delta \hat{J}^2$, where $\hat{J}$ is the total angular momentum operator with eigenstates $|J,m_J\rangle$. The energies of these eigenstates are lifted by intrinsic hyperfine interactions, as well as tunable extrinsic effects resulting from applied electric and magnetic fields and incident laser light. These extrinsic effects set the molecules' quantization axes and enable a direct modulation of the rotational energy levels, and hence the two sub-lattices.

Focusing on the four lowest-energy rotational eigenstates, we define two distinct hard-core bosonic degrees of freedom. On the $A$-sub-lattice we utilize $\{ |0_A \rangle = |0,0\rangle_A, |1_A \rangle = |1,0\rangle_A \}$, while on the $B$-sub-lattice we utilize $\{ |0_B \rangle = |0,0\rangle_B, |1_B \rangle = |1,1\rangle_B \}$, as illustrated in Fig.~\ref{fig:expt}. These hard-core bosons interact with one another via a dipolar interaction, which gives rise to the effective hoppings:
\vspace{-.3cm}
\begin{equation}\label{ts}
\begin{split}
t^{AA}_{\mathbf{r}} & = - C^{AA} \,  \frac{3\cos^2(\theta) - 1}{R^3} \\
t^{BB}_{\mathbf{r}} & = C^{BB}  \, \frac{3\cos^2(\theta) - 1}{R^3} \\
t^{AB}_{\mathbf{r}} = \left( t^{BA}_{-\mathbf{r}} \right)^* & = -C^{AB} \, \frac{\cos(\theta)\sin(\theta)}{R^3} \, e^{i\phi} \\
\end{split}
\end{equation}
where $\left\{R,\theta,\phi\right\}$ defines the separation of molecules in spherical polar co-ordinates, and $C^{AA}, C^{BB}$, and $C^{AB}$ are positive constants. Details are provided in the companion manuscript~\cite{schuster2021floquet}. This particular choice of rotational states ensures that the inter-sub-lattice hopping $t^{AB}_{\mathbf{r}}$ is induced solely by the $\Delta m_J = -1$ term, which immediately gives rise to a hopping phase $\propto e^{i\phi}$~\cite{fnQWZ}. This choice, motivated by the model of Ref.~\cite{moore2007topological}, locks the intra-sub-lattice components of the Hamiltonian $n_{x,y}(\mathbf{k})$ to the momenta $k_x,k_y$. As illustrated in the companion manuscript~\cite{schuster2021floquet}, this locking naturally achieves the Hopf requirement that all Bloch sphere pre-images link.

We further enhance the relative strength of next-nearest neighbor hopping with a simple Floquet engineering strategy. The basic principle is that, by periodically modulating the on-site chemical potentials $\mu^\alpha_\mathbf{v}\left(t\right)$ inhomogeneously at frequencies, $\hbar \Omega$, significantly higher than the energy of the dipolar interaction, the time-averaged behavior emulates that of a different time-independent Hamiltonian. In this effective Hamiltonian, sites that oscillate out-of-phase with one other will have the hopping between them suppressed, while hoppings between sites oscillating in-phase remain unaffected. Although the Floquet modulation $\mu^{\alpha}_{\mathbf{v}}\left(t\right)$ necessarily varies with the lattice site $\mathbf{v}$, we choose it such that effective hoppings remain translationally-invariant. Specifically, we take the Floquet modulation to be a checkerboard pattern in the $xy$-plane, such that next-nearest-neighbor hoppings (even $r_x+r_y$, in-phase) are enhanced relative to nearest-neighbour hoppings (odd $r_x+r_y$, out-of-phase).

\FigEdgeStates

Additionally, although the slow decay of the $1/R^3$ dipole-dipole interaction is helpful in establishing the next-nearest neighbor interactions in the $xy$-plane that are necessary to realize the HI, our numerical studies indicate that the same interactions cause unnecessary long-range couplings in the $z$-direction. To address this, we utilize an additional, second patterning of the previous Floquet engineering strategy, which truncates the dipolar interaction to effectively nearest-neighbor in the $z$-direction~\cite{lee_floquet_2016}. This patterning is guaranteed to operate independently of the previous Floquet engineering patterning if their modulation frequencies are well-separated in scale; we verify this quantitatively in the companion manuscript~\cite{schuster2021floquet}.

By this process we are able to identify parameters in Eq.~\eqref{eq:Htb} that realize the Hopf insulating phase, $h = 1$, with band gaps as large as $E_g \gtrsim 0.26 \, t_{\text{nn}}$ (in units of the nearest-neighbor hopping), as well as gapless transitions between the Hopf and trivial insulating phases~\cite{schuster2021floquet}. Utilizing the $\Delta m_J = +1$ component of the dipolar interaction (as opposed to the $\Delta m_J = -1$ component)  leads instead to the phase $h=-1$; higher linking numbers are in principle possible, but require an even more delicate structure in $k$-space.

In Fig.~\ref{fig:edge_states} we show the band structure found by exact diagonalization of the dipolar Hamiltonian [Eqs.~(\ref{eq:Htb}) and ~(\ref{ts})] after applying our Floquet engineering strategy~\cite{schuster2021floquet}. 
We assume periodic boundary conditions in the $y$- and $z$-directions (crystal momenta $k_y$ and $k_z$ are therefore good quantum numbers) but a finite length in the $x$-direction. 
We also truncate the hopping range to $|\mathbf{r}|\le 8$ sites for numerical feasibility; increasing the truncation range does not qualitatively affect the results. 
Fig.~\ref{fig:edge_states}(b) shows the result of a smooth adiabatic termination over twenty lattice sites. The bulk (black) is gapped, but the edges (red and blue) host conducting states, which we found to be stable for any sizable bulk gap. This is the Hopf insulator: the adiabatic termination approximately preserves translational-invariance, leading to the survival of the two-band picture. Since the Hopf invariant is trivial outside the system and unity inside, gapless edge states result at the interface.  

Fig.~\ref{fig:edge_states}(c) shows the result of an abrupt termination of the edge. Lacking adiabaticity, the band picture is expected to break down; since the HI exists only for two-band models, we would then not expect topologically-protected edge states. Remarkably, however, edge states are again present. In fact, a serendipitous choice of edge-termination plane $\left(100\right)$ has lead to the bulk $\mathcal{J}$ symmetry surviving at the edge, and these edge states are a manifestation of the resulting strong $\mathbb{Z}_2$ invariant (which does not require a two-band model). To see this `accidental' symmetry, note that open boundary conditions are equivalent to an infinite potential barrier $\hat{H}_{\text{edge}} =  \rho \hat{\sigma}^z \delta_x$, $\rho \rightarrow \infty$ at the system's edge, where $\hat{\sigma}_z$ acts on the sub-lattice degrees of freedom. In momentum space, this corresponds to real couplings between different $k_x$, $\hat{H}^{\b{k},\b{k}'}_{\text{edge}} = \rho \hat{\sigma}^z \delta_{k_y,k'_y} \delta_{k_z,k'_z}$, which is easily seen to obey Eq.~(\ref{eq:Cenke}). Nearly any perturbation to naive open boundary conditions -- for instance a small potential $\gamma \, \hat{\sigma}^z \delta_{x-1}$ on the site nearest the edge -- breaks the $\mathcal{J}$ symmetry and gaps the edge states [Fig.~\ref{fig:edge_states}(d)]. We predict that this same mechanism is responsible for stabilizing the edge states at sharp boundaries observed in previous numerical studies of the HI~\cite{moore_topological_2008,deng_hopf_2013}. All of these edge mode structures can be probed experimentally via molecular gas microscopy~\cite{marti2018imaging,covey2018approach} by exciting individual edge spins and observing the extent to which the excitation remains localized on the edge~\cite{schuster2021floquet}. 

Before concluding, we detail the separation of scales required for Eqs.~(\ref{eq:Htb}) and~(\ref{ts}) to govern the low-energy dynamics of the polar molecular system. First, we work in the natural experimental regime where the dipolar interaction strength is significantly smaller than the energy splittings between the rotational states within the $J=1$ manifold. The external fields should be tuned such that the splitting between the $|0_A \rangle$ and $ |1_A \rangle$ states is resonant with the $|0_B\rangle$ and $ |1_B\rangle$ states, and far detuned from all other rotational transitions. Conservation of energy then dictates that the dipolar interaction can only induce transitions within our prescribed hard-core bosonic doublets. Details on how this level scheme can be precisely realized in the specific case of polar molecular quantum simulation based upon ${}^{40}$K$^{87}$Rb can be found in the companion manuscript~\cite{schuster2021floquet}. Here, we note only that the orientation of the spins is fixed via applied fields oriented in the $z$-direction, and that the degeneracy between the $|1,0\rangle$ and $|1,1\rangle$ states, as well as the sub-lattice symmetry between the $A$ and $B$ planes, is broken by using \emph{different} intensities of light to form each sub-lattice. Our scheme naturally leads to a separation of energy scales $t \ll \delta \ll \Delta$, where $t$ is the dipolar interaction strength ($\sim \!\! 100$ Hz), $\delta$ is the splitting within the $J=1$ manifold ($\sim \!\! 5$ kHz), and $\Delta$ is the splitting between the $J=0$ and $J=1$ sectors ($\sim \!\! 2$ GHz).
 
There has recently been a burst of theoretical interest in Hopf insulators and their possible extensions, including non-hermitian generalizations~\cite{he2020non}, the survival of topology under quantum quenches~\cite{hu2020quench}, crystal symmetries~\cite{liu_symmetry_2017, alexandradinata2019actually}, and generalizations to periodically-driven Floquet systems~\cite{schuster2019floquet,he2019three}. These ideas motivate the possibility of experimentally realizing the Hopf insulator phase, which would allow one to test the above predictions, and, more tantalizingly, could probe regimes of Hopf insulating physics that are much harder for theory to handle. For instance, it remains an open question as to whether any interacting extension of the Hopf insulator exists.
The protocol outlined here and detailed further in the companion manuscript~\cite{schuster2021floquet} makes use of the high tunability and intricate real- and momentum-space structures afforded by recent advances in the manipulation of interacting dipolar molecules~\cite{valtolina2020dipolar,matsuda2020resonant}. Looking forward, the same approach suggests many promising avenues for realizing other exotic states presently residing at the forefront of theory~\cite{qi2006topological,agarwala2019topological,wan2011topological,grushin2014floquet,regnault2011fractional,zhou2013unconventional}. %
 
%Looking forward, our proposal suggests a number of intriguing directions ranging from many-body generalizations of the Hopf insulator to novel realizations of hybrid Hopf-Chern insulators \cite{kennedy_topological_2016}.

\emph{Acknowledgments}---We gratefully acknowledge the insights of and discussions with Dong-Ling Deng, Luming Duan, Vincent Liu, Kang-Kuen Ni, and Ashvin Vishwanath. This work was supported by the AFOSR MURI program (FA9550-21-1-0069), the DARPA DRINQS program (Grant No. D18AC00033), NIST, the David and Lucile Packard foundation, the W. M. Keck foundation, and the Alfred P. Sloan foundation. T.S. acknowledges support from the National Science Foundation Graduate Research Fellowship Program under Grant No. DGE 1752814. F. F. acknowledges support from a Lindemann Trust Fellowship of the English Speaking Union, and the Astor Junior Research Fellowship of New College, Oxford. Work at Temple University is supported by ARO Grant No. W911NF-17-1-0563, AFOSR Grant No. FA9550-21-1-0153, and NSF Grant No. 1908634. 

\bibliographystyle{apsrev4-1}
\bibliography{refs_DipolarHopf,refs_SvetlanaMing} 

\end{document}